# Phase Walk Analysis of Leptokurtic Time Series

Korbinian Schreiber,[1, a)] Heike I. Modest,[2, b)] and Christoph Räth[2, c)]
[1)] *Heidelberg University, Im Neuenheimer Feld 227, 69120 Heidelberg*
[2)] *Deutsches Zentrum für Luft- und Raumfahrt, Institut für Materialphysik im Weltraum, Münchner Str. 20, 82234 Wessling, Germany*

(Dated: 4 June 2018)

The Fourier phase information play a key role for the quantified description of nonlinear data. We present a novel tool for time series analysis that identifies nonlinearities by sensitively detecting correlations among the Fourier phases. The method, being called *phase walk analysis*, is based on well established measures from random walk analysis, which are now applied to the unwrapped Fourier phases of time series. We provide an analytical description of its functionality and demonstrate its capabilities on systematically controlled leptokurtic noise. Hereby, we investigate the properties of leptokurtic time series and their influence on the Fourier phases of time series. The phase walk analysis is applied to measured and simulated intermittent time series, whose probability density distribution are approximated by power laws. We use the day-to-day returns of the Dow-Jones industrial average, a synthetic time series with tailored nonlinearities mimicing the power law behavior of the Dow-Jones and the acceleration of wind at an Atlantic offshore site. Testing for nonlinearities by means of surrogates shows that the new method yields strong significances for nonlinear behavior. Due to the drastically decreased computing time as compared to embedding space methods the number of surrogate realizations can be increased by orders of magnitude. Thereby, the probability distribution of the test statistics can very accurately be derived and parametrized which allows for much more precise tests on nonlinearities.

A very clear and general definition of nonlinearity in data sets can be obtained from their representation in Fourier space: From the Wiener-Khintchine theorem on the one hand and the bijectivity of the Fourier transformation on the other hand it follows that the linear information is entirely represented by the Fourier amplitudes. Hence, *all* nonlinear information is contained solely in the Fourier phases. This reasoning is also fundamental for the development of the method of surrogates for testing for nonlinearities, where surrogates are generated by randomizing the Fourier phases and by preserving only the linear correlations. Yet, the direct study of the Fourier phases has so far attracted only little attention. Here, we present a novel method to quantify the phase information. In close analogy to the well-established methods for random walk analysis, we propose the phase walk analysis as a way to quantify the phase information. We apply it to the analysis of nonlinearities in intermittent, leptokurtic time series like the Dow Jones day-to-day return, wind data and synthetic leptokurtic data and outline the capabilities of the novel approach for detecting and assessing nonlinearities.

---

[a)] Electronic mail: k.schreiber@kip.uni-heidelberg.de
[b)] Electronic mail: heike.modest@dlr.de
[c)] Electronic mail: christoph.raeth@dlr.de

## I. INTRODUCTION

Detecting nonlinear features in time series is often accomplished by comparing the result of a suitable nonlinear measure to the corresponding results from a surrogate data set. Commonly used nonlinear measures are for example the Lyapunov exponent,[1–4] errors in nonlinear prediction[5,6] or multifractal dimension estimates.[7] All of these methods have to be applied to the higher dimensional representations of the time series known as attractors that are mostly obtained by delay-coordinate-embedding.[8,9] A different attempt to identify nonlinear properties has been made in the analysis of the Cosmic Microwave Background. Here, it turned out very beneficial to search for correlations among the phases of the spherical harmonics, referred to as Fourier phases, to identify deviations from Gaussianity.[10–15] Also in the evolution of cosmic large scale structures an increase of such phase correlations has been observed.[16–19] A similar approach helped to uncover nonlinearities induced by certain surrogate generating algorithms.[20] Moreover, fundamental scaling properties of highly nonlinear financial time series have exactly been reproduced by imposing a set of correlations on the Fourier phases of Gaussian white noise.[21]

For a comprehensive description of nonlinear data it is helpful to include the quantification of Fourier phase information. It was already stated by Ruelle and Eckmann in their seminal review paper on chaos and strange attractors[22] that *"[...] the analysis of the chaotic motions themselves does not benefit much from the power spectra, because (being squares of absolute values) they lose phase information, which is essential for the understand-* ing of what happens on a strange attractor". Yet, little effort has so far been put into quantifying and thus understanding the information contained in the phases of nonlinear time series.[23–25]

By *nonlinear* we refer to all features that are not captured by (linear) autoregressive-moving-average-models (ARMA).[26,27] The ARMA parameters comprise the linear properties of the time series data and are fully represented by a set of coefficients that can bijectively be mapped onto the autocorrelation coefficients by the Yule-Walker equations.[28,29] The autocorrelation function – consequently carrying all linear traits – can in turn bijectively be mapped onto the power spectrum by a Fourier transform, as described by the Wiener-Khinchin-Theorem.[30] Since the power spectrum is defined as the squared amplitudes of the Fourier coefficients, all information left to fully reconstruct the original time series comes from the phases of the Fourier coefficients. The nonlinear properties of a times series are therefore fully represented by its Fourier phases. In the linear case these phases are uniformly and identically distributed whereas in the nonlinear case they show various kinds of correlations among them.

Nonlinearity thus refers to all those features in time series that are not captured by the power spectrum. A clear categorization of the different sorts of nonlinearities is still far from being established but would allow for a more accurate model selection.

A crucial distinction is made between static and dynamic nonlinearities.[31] The former are deviations from a Gaussian distribution of the data points in real space which can already be induced by a static nonlinear transformation. The latter refer to nonlinear correlations in the time series induced by the dynamics of the system. It has been argued[20] that due to shortcomings of existing surrogate generating algorithms one has to test for static and dynamic nonlinearities separately. We follow this reasoning and outline on carefully selected examples how a test statistics based on the information contained in the Fourier phases can give complementary insights into static nonlinearities.

Our ansatz for tracing anomalies in the Fourier phases is based on applying random walk statistics to the *unwrapped*[32] Fourier phases. Instead of considering a path of random steps in time we define random steps between Fourier phases following the phase index.

We present an analytic description that models the effects of leptokurtosis as a simple form of a static nonlinearity on the unwrapped Fourier phases. Then we introduce a methodology that captures these phase anomalies and validate the newly introduced method with a statistical analysis of random distributions with defined kurtosis. Finally, we apply the phase walk method to empirical and synthetic data. The test examples include a financial time series, a time series of wind velocities in a turbulent environment and an artificial time series with tailored nonlinearities. All of these are known to carry leptokurtic probability distributions of the observ-ables and dynamic features as well.[21,33–37] We conclude with outlining the performance and capabilities of the novel approach compared to other techniques in nonlinear time series analysis.

## II. METHODS

### A. Fourier transformation and phase unwrapping

The Fourier transform of a discrete time series $g(t)$ is given by

$$G(k) = FT(g(t)) = \frac{1}{N}\sum_{t=0}^{N-1} g(t)e^{i2\pi kt/N} = |G(k)|e^{i\phi(k)}. \tag{1}$$

Here, $|G(k)|$ are the moduli or amplitudes of the Fourier coefficients $G(k)$, $\phi(k)$ are the corresponding phases with values ranging in the interval $I_\phi = ]-\pi,\pi]$ and $N$ is the number of time steps. An intrinsic feature of a numerically computed discrete Fourier transform is that all phases are *wrapped* onto the interval $I_\phi$. Therefore, possible trends in the course of the phase sequence cannot be easily recognized. Linear trends for example lead to a sawtooth progression of the $\phi_k$ (see Fig. 1). To overcome this problem, a technique known as phase tracking or unwrapping is applied and resolves the true phases by reinterpreting the differences between two consecutive phases. In this paper we use a simple algorithm, introduced by Itoh.[32] If the phase difference or "rotation" $\Delta\phi(k) = \phi(k+1) - \phi(k)$ exceeds $\pi$ – which is considered as counterclockwise rotation – the rotation is reinterpreted as being clockwise by subtracting $2\pi$. Likewise, when $\Delta\phi(k)$ is less than $-\pi$ (clockwise), it is reinterpreted as counterclockwise by adding $2\pi$. To construct the differences of the unwrapped phases $\Delta\phi'(k)$, the following rule can thus be applied:

$$\Delta\phi'(k) = \begin{cases} \Delta\phi(k) - 2\pi & \text{if } \Delta\phi(k) > +\pi \\ \Delta\phi(k) + 2\pi & \text{if } \Delta\phi(k) \leq -\pi \\ \Delta\phi(k) & \text{else} \end{cases} \tag{2}$$

The unwrapped phases are then obtained by a cumulative sum over the differences with $\phi'(0) = \phi(0) = 0$:

$$\phi'(k) = \sum_{i=0}^{k-1} \Delta\phi'(i) \tag{3}$$

The procedure is illustrated for an artificially constructed series of phases in Fig. 1.

Noisy data, which is by far the most common case in time series analysis, can induce so called *fake wraps*.[32,38] This motivated the development of more sophisticated unwrapping algorithms (see for example Ghiglia et al.[39]). Most of them though assume the Nyquist criterion[40] because aliasing might lead to wrong or discontinuous phase differences. Additionally, they require a relatively high signal-to-noise ratio for the phase transitions to be smooth. Both is not in general the case for experimental data and in particular, it does not hold true for the data investigated here. We will therefore not gain advantage by using these kinds of algorithms and stick with the presented variant.

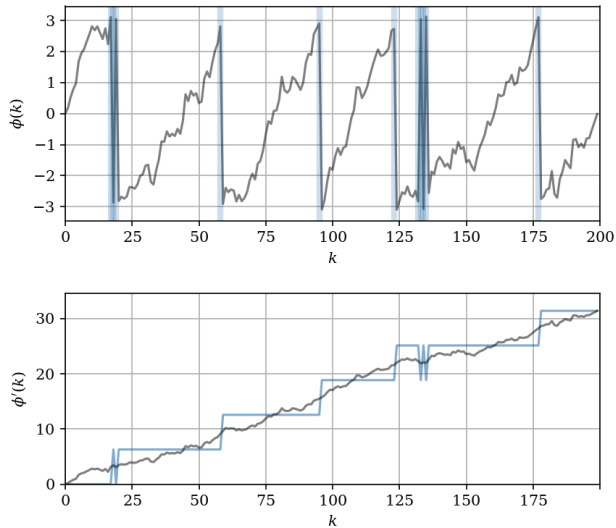

FIG. 1. Top: An artificially constructed series of Fourier phases with an overall positive linear trend. The phases are initially constrained to the interval $I_\phi =\,]-\pi,\pi]$. The blue vertical lines indicate differences between two consecutive phase values with $|\phi(k+1) - \phi(k)| > \pi$. Bottom: A reconstruction of the unwrapped phase series (gray) according to Itoh's algorithm.[32] Here, the blue curve represents the accumulated correction by the unwrapping algorithm.

Note here that, due to the periodicity of the phases, the unwrapping procedure leaves the Fourier phases in a modified state which, however, has no consequence for the Fourier representation at all. A back transformation of a Fourier representation of a time series with wrapped phases would yield the very same time series as of one with unwrapped phases.

### B. Outliers and Fourier phases

Uncorrelated Gaussian or more generally mesokurtic time series have a set of random, uncorrelated and uniformly distributed Fourier phases. A time series of uncorrelated, leptokurtic noise $\xi(t)$ can be expressed as a superposition of a mesokurtic noise floor $\eta(t)$ and defined shifts of specific data points:

$$\xi(t) = \eta(t) + \sum_{j=0}^{P-1} a_j \delta(t - \tau_j). \quad (4)$$

Here $a_j$ is the amount of shift of the $j^{th}$ data point at the temporal position $\tau_j$. We will call these points *outliers* from now on. $P$ is the number of outliers and the *delta-function* is defined as $\delta(t - \tau_j) = 1$ if $t = \tau_j$ and 0 else. The discrete Fourier transform $\Xi(k)$ of $\xi(t)$ then simply becomes

$$\Xi(k) = H(k) + \sum_{j=0}^{P-1} a_j e^{i2\pi \frac{\tau_j}{N} k} = H(k) + \sum_{j=0}^{P-1} A_j(k). \quad (5)$$

$H(k)$ is the discrete Fourier transform of $\eta(t)$. We see that also the Fourier representation of $\xi(t)$ is a superposition of $H(k)$ and the complex numbers $A_j(k)$ that represent vectors of length $a_j$ and angle

$$\phi_j(k) = \frac{2\pi \tau_j}{N} k \propto k \quad (6)$$

in the complex plane. For all $\tau_j \neq 0$, the angle increases with $k$. For $\tau_j = 0$, also $\phi_j$ becomes zero. To illustrate the effect of those $A_j(k)$, we first consider the simplest case of only one dominating outlier, represented in Fourier space by $A_0(k)$. The phase $\Phi(k)$ of $\Xi(k)$ then becomes the angle of the sum of the two vectors $A_0(k)$ and $H(k)$ in the complex plane. To obtain a strong leptokurtic effect from a single outlier, we require that it is shifted by an amount $a_0$ that is significantly larger than the average time series amplitude. In turn, $|A_0(k)| = a_0$ will strongly dominate over $|H(k)|$. Hence, $\Phi(k)$ will be $\phi_j(k)$ plus a relatively small random fluctuation around zero, since the direction, or phase, of $H(k)$ is randomly and uniformly distributed for Gaussian noise. As $\phi_j(k) \propto k$, also $\Phi(k)$ grows linearly with $k$. Thus, the unwrapped phases increase linearly, where the slope is determined by the position of the single outlier. The effect of one outlier on the (unwrapped) phases is illustrated in Fig. 2.

If there is more than one outlier, a more complicated picture in the complex plane arises. The noise floor $H(k)$ can still be regarded as random numbers with a relatively small magnitude compared to the $a_j$ and a uniformly distributed angle. The $A_j(k)$ however rotate with a slope of $\dot\phi_j = \frac{d\phi_j}{dk} = \frac{2\pi\tau_j}{N}$ as $k$ increases. Note, that also the slope increases as the temporal position $\tau_j$ of the spike moves from the beginning of the time series to its end. In the complex plane, this corresponds to a chain of rotating vectors with a small random fluctuation, given by $H(k)$. If the number of outliers with roughly equal magnitude is large (this would not meet the assumption of leptokurtosis), this results in a complex superposition of cycles and epicycles. But if there is still a limited number of dominating outliers an overall trend in the rotation of $\Xi(k)$ can be recognized. This is the case for leptokurtic and especially scale-free distributions. Another point to mention is that the direction of the vector $\Xi(k)$ can also be essentially steered into one particular direction, if the $A_j(k)$ rotate nearly coherently with $k$. This happens, if many $\tau_j$ share roughly equal values, corresponding to a cumulation of outliers and hence, to a burst event in the time series. For financial data this occurs with volatility clustering.[33,41]

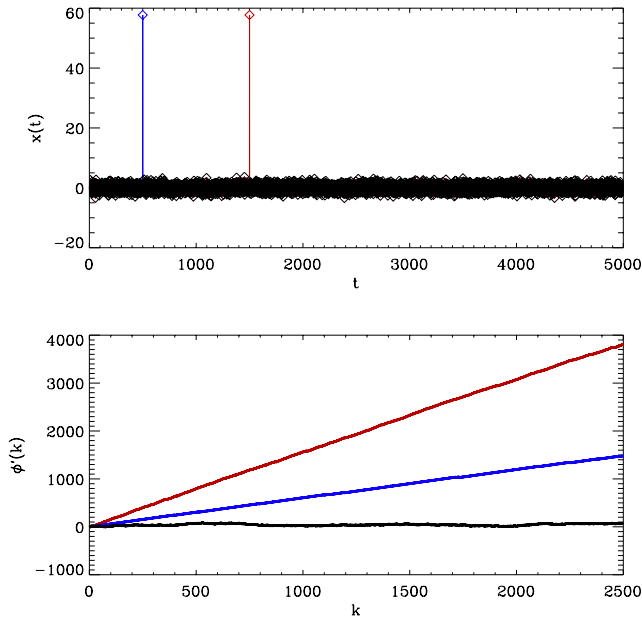

FIG. 2. Top: Gaussian noise with one outlier at $t = 500$ (blue) and $t = 1500$ (red). Bottom: Unwrapped phases for the Gaussian noise (black) and the time series with the outlier at $t = 500$ (blue) and $t = 1500$ (red).

### C. Phase Walks

The difference between two consecutive unwrapped phases is again constrained to an interval $I_{\Delta\phi'(k)} = I_{\phi'(k+1)-\phi'(k)} = [-\pi, \pi]$ and distributed according to a probability density function $P(\Delta\phi'(k) = x)$. Writing the unwrapped phases as $\phi'(k+1) \equiv \phi'(k) + \Delta\phi'(k)$ resembles the exact form of a one-dimensional random walk that depends on the phase index $k$ instead of the time $t$. This feature allows us to apply well established methods from random walk analysis to the unwrapped phases in order to test them for anomalies like trends or periodicities. Following this perspective, we also refer to the unwrapped phases by the term *phase walk*. To detect anomalies in phase walks with quantifiable significance, we need to formulate the null hypothesis, which we first choose to be the phase walk of Gaussian, uncorrelated (white) noise $\{\eta(t)\}_{t=0}^{N-1}$. The Fourier phases of this kind of noise are identically and independently distributed: $P(\phi(k) = x) = P_{\phi(k)}(x) = 1/2\pi$ if $x \in [-\pi, \pi]$ and 0 else. It can be shown that the same distribution describes the differences between the steps of the unwrapped phases in this case. Thus,

$$P_{\Delta\phi'(k)}(x) = P_{\phi(k)}(x) = \begin{cases} \frac{1}{2\pi} & \text{if } x \in [-\pi, \pi] \\ 0 & \text{else} \end{cases}. \quad (7)$$

Furthermore, $\phi(0) = \phi'(0) = 0$, as we deal with a real signal $\eta(t)$. The variance corresponding to the single step distribution in Eq. (7) is then given by $\sigma_1^2 = \pi^2/3$. The index 1 indicates the number of steps, or *lag interval*, between phase $\phi'(k)$ and $\phi'(k+1)$. Due to the central limit theorem the distribution function for a lag interval of $k$ steps is Gaussian with a variance $\sigma_k^2 = k\sigma_1^2$. In Fig. 3 we show 11 examples of random walks and phase walks as well as the corresponding probability distributions of the data points at $k = 999$. If the phase walks behave like this, we say that they fulfill the *random walk hypothesis* (RWH). While this holds true for various types of noise, like e.g. colored noise or Poisson noise, it does not in general. To detect and quantify deviations we introduce a slightly adapted version of the variance ratio test.

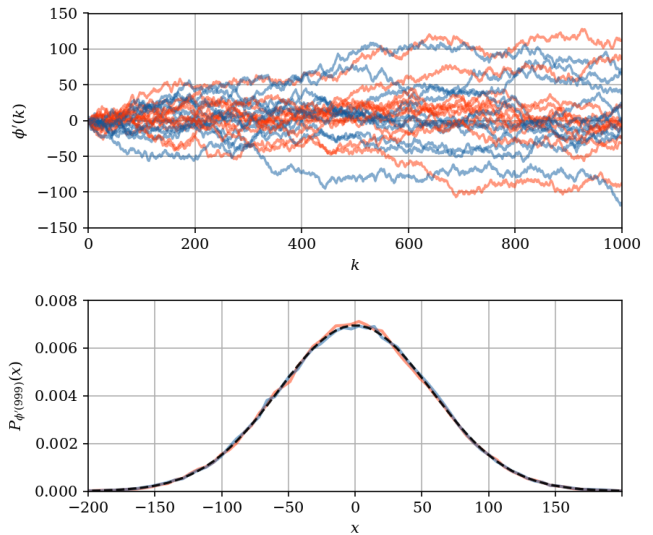

FIG. 3. Top: 11 random walks (red) with 999 uniformly distributed steps according to Eq. (7) and 11 phase walks (blue) extracted from Gaussian noise. Bottom: The empirical distributions of the last step $k = 999$ of constructed random walks (red) and extracted phase walks (blue). Both are based on $10^5$ realizations. Additionally, a Gaussian function with variance $\sigma_{999}^2 = 999 \cdot \pi^2/3$. The phase walks of Gaussian noise behave exactly like the ideal random walks with uniform step distribution.

### D. Standard deviation ratio test (SRT)

Variance ratio or standard deviation ratio tests are applied to decide whether a given time series follows the dynamics an ideal random walk. These tests compare a variance determined from the tested data to an ideal variance $\sigma_\kappa$. If a phase walk fulfills the random walk hypothesis, the ideal variance of the distribution of the differences of two data points separated by a distance $\kappa$ can be calculated as[42–44]

$$\sigma_\kappa^2 = \kappa \cdot \sigma_1^2 = \frac{\pi^2}{3}\kappa. \quad (8)$$

One can compare this variance to an estimate of the variance of the tested phase walk obtained by averaging.[45]

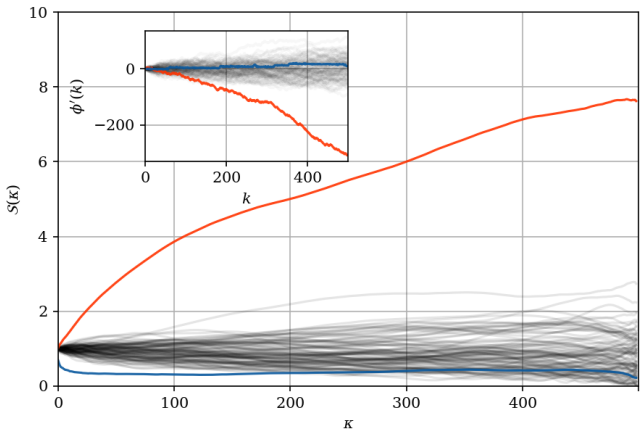

FIG. 4. $\mathcal{S}(\kappa)$ for 100 ordinary phase walks (black), for one phase walk with a significant trend, and for one phase walk that is centered (blue). The embedded plot shows the corresponding unwrapped phase walks. For all the ordinary phase walks, $\mathcal{S}(\kappa)$ takes on values between 0 and 3, while the distribution stays pretty tight for small values of $\kappa$. On the other hand, $\mathcal{S}(\kappa)$ increases rapidly for the phase walk with a trend and it decreases rapidly for the centered phase walk.

Taking the square root of this ratio leaves us with a ratio $\mathcal{S}(\kappa)$ of standard deviations

$$\mathcal{S}(\kappa) = \sqrt{\frac{\langle [\phi'(k+\kappa) - \phi'(k)]^2 \rangle_k}{\sigma_\kappa^2}}$$
$$= \sqrt{\frac{3 \cdot \sum_{k=0}^{K-\kappa-1} [\phi'(k+\kappa) - \phi'(k)]^2}{\pi^2 \kappa (K-\kappa)}}. \quad (9)$$

$K$ is the total number of phases and the index difference $\kappa$ is also referred to as *lag phase*. Test results that significantly deviate from the RWH will either imply correlations between the steps, deviations from the distribution of the steps $p_{\Delta\phi'(k)}(x)$, or both. If $\mathcal{S}(\kappa)$ stays below three, the phase walk is likely to fulfill the RWH. If $\mathcal{S}(\kappa) \ll 1$ for a wide range of $\kappa$, the phase walk is likely to be centered. Finally, if $\mathcal{S}(\kappa) > 3$, the phase walk is likely to have a trend. For a statistical analysis of a large number of test results, it is convenient to evaluate $\mathcal{S}(\kappa)$ only for one fixed value of $\kappa$. For phase walks with trends, $\mathcal{S}(\kappa)$ becomes most significant for large values $\kappa$ (see Fig. 4). Since trends in the phase walks are what we expect for leptokurtic time series, we maximize $\kappa$ by approximately setting it to the maximum phase index in the following examples.

We like to conclude with a brief summary of the whole test procedure:

1. Compute the Fourier transform of the time series

2. Unwrap the Fourier phases according to Eq. (2) & Eq. (3)

3. Compute $\mathcal{S}(\kappa)$ according to Eq. (9)

4. Plot and interpret the result

5. For a statistical analysis, collect a large number of results for a suited, and fixed value of $\kappa$

## III. EXAMPLES

### A. Artificial leptokurtic time series

To show empirically that static nonlinearities are directly related to phase correlations we constructed noise time series with leptokurtic data point distributions, by drawing 20,000 random variables from an adjusted Pearson type VII distribution (Student's t distribution):

$$p(x|\gamma_2) = \frac{\Gamma(\frac{5}{2} + \frac{3}{\gamma_2})}{\sqrt{2\pi(1+\frac{3}{\gamma_2})}\Gamma(2+\frac{3}{\gamma_2})}(1 + \frac{x^2}{2+\frac{6}{\gamma_2}})^{-\frac{5}{2}-\frac{3}{\gamma_2}}. \quad (10)$$

This provides the option to control the kurtosis parameter $\gamma_2$ of the distribution if $\gamma_2 \geq 0$. For the platykurtic samples with $\gamma_2 = -6/5$ we used a uniform distribution instead. In both cases, the standard deviation is normalized to 1. Fig. 5 shows the distributions of $\mathcal{S}(\kappa)$ from the standard deviation ratio test at a fixed lag phase $\kappa = 9,800$ for various values of $\gamma_2$. For each $\gamma_2$ we constructed 100,000 different random time series. It becomes clear that $P_{\mathcal{S}(\kappa)}(x)$ gets wider monotoniously as $\gamma_2$ increases. Large values of $S(\kappa)$ – indicating trends in the phase walks – become more likely, which confirms a statistical influence of heavy tails on the Fourier phases as expected in Sec. II B.

### B. Empirical Data

To demonstrate the test's significance for dynamic nonlinearities, three time series known to originate from highly nonlinear processes were selected: First, the logarithmic daily returns of the Dow-Jones industrial average (DJ),[46] reaching from 26$^{\text{th}}$ of May 1896 until 23$^{\text{rd}}$ of October 2014, resulting in $N = 32222$ time steps.
The second time series has been synthesised to match all scaling properties of the data point distribution of the DJ data set and to reproduce all static nonlinearities of the original DJ time series. This was achieved by imposing a set of six linear correlations on the Fourier phases as proposed by Räth and Laut.[21] While the original time series has $15 \times 10^5$ time steps, only a shorter version that is cropped to the exact length $N$ of DJ is used in the current analysis. This synthetic time series is called tailored nonlinearity time series (TN). We demonstrate that the phase walk analysis can find differences between this time series and its prototype DJ.
Third, a time series of wind velocities collected at an Atlantic offshore wind turbine[47] serves as a leptokurtic example that also shows some non-negligible linear properties. To make the third time series comparable to the previous two examples, it is cropped to the length of DJ and furthermore, *detrended* by taking the differences between the time steps. As a result, the time series reaches from $1^{\text{st}}$ of January 2004, 00:10 until $10^{\text{th}}$ of July 2004, 18:20. The difference between two consecutive velocities is the change in velocities and hence, the time series describes the wind acceleration (WA).

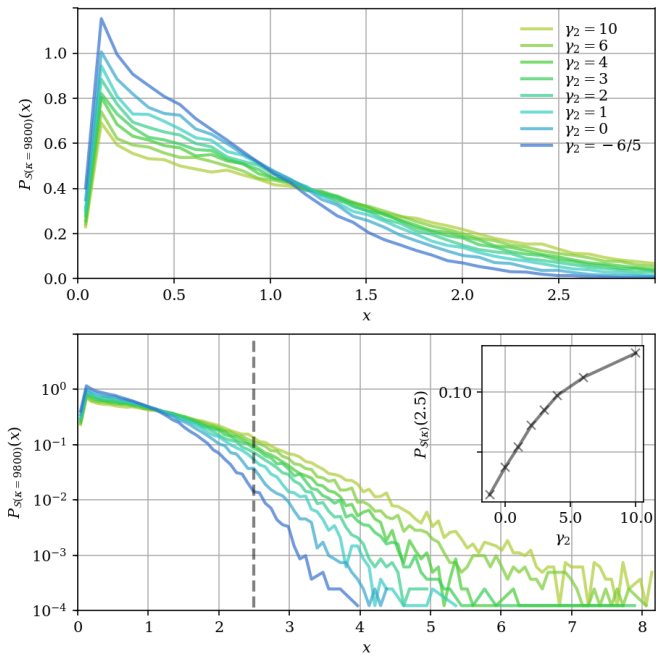

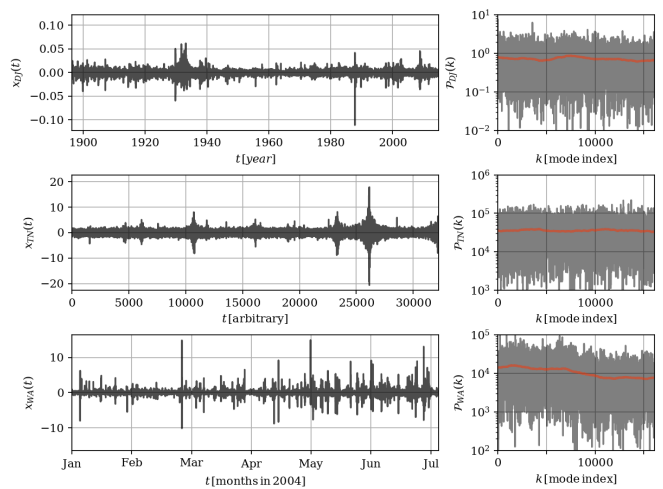

FIG. 5. Top: Empirical probability distributions of the SRT results for noise with varying kurtosis. Bottom: The same distributions on a logarithmic scale. The small embedded window shows a slice (indicated by the dashed gray line) through the distributions at $x = 2.5$. The probability density at $x = 2.5$ increases monotonically with $\gamma_2$. Each distribution is based on $100,000$ randomly generated time series with $20,000$ time steps each.

FIG. 6. Top: Logarithmic daily returns of the Dow Jones Industrial Average from $26^{\text{th}}$ of May 1896 until $23^{\text{rd}}$ of October 2014. Mid: Artificial time series with tailored nonlinearities. Bottom: Wind acceleration measured at an Atlantic offshore wind turbine from January, $1^{\text{st}}$, 2004, 00:10 until Juli, $10^{\text{th}}$, 2004, 18:20 with a sampling period of 10 minutes. In the right column the corresponding power spectrum (gray) and its trend (red) obtained by averaging over 2000 neighbors for each Fourier mode is displayed.

All three time series and their power spectra are shown in Fig. 6. The right column of Fig. 6 suggests flat power spectra for DJ and TN and thus no nameable linear features. On the other hand, the spectrum of WA drops slightly for higher frequencies. $\mathcal{S}(\kappa)$ will capture the static nonlinearities – related only to the non-Gaussian shape of the distributions in real space – as well as all other possible nonlinear contributions. To separate these, we perform a surrogate assisted statistical evaluation of the test results.[31] The surrogates are generated using two different methods:

1. We randomize or *shuffle* the order of all the time steps to destroy any temporal correlation. The surrogates therefore preserve the data point distribution in real space, but loose the linear properties. Since DJ and TN already have white power spectra, their linear properties will be reproduced statistically though. As Dolan and Spano[48] argue, a non-exact replication of the original time series' power spectrum may even result in better null hypothesis.

2. We create iterated amplitude adjusted Fourier transform (IAAFT) surrogates.[31,49] This algorithm preserves both, the power spectra as well as the data point distribution of the original time series. For WA the IAAFT surrogates are more favorable than the shuffled variant, as it contains some relevant features in the power spectrum.

For each time series, we created $10^5$ surrogates using both methods respectively. These surrogates serve as the null hypothesis for the test.

Fig. 7 shows the empirical probability distributions of $\mathcal{S}(\kappa)$ for the different time series and algorithms and additionally the results for the original data. We choose the maximal lag phase by setting $\kappa$ to almost the total number of phases $K$: $\kappa = K - 2 = 16108$. As mentioned before, TN has been tailored to exactly match the data point distribution of its prototype time series DJ and hence, it is not surprising to find their null distributions very similarly lying upon each other. The results for the actual time series on the other hand differ strongly ($\mathcal{S}_{DJ} = 20.0$ and $\mathcal{S}_{TN} = 32.9$). The SRT method indicates even greater significance for nonlinearities in TN. This implies that although the static nonlinearities have been reproduced very accurately, the dynamic nonlinearities of both data sets still deviate by a large amount. The test result distributions associated with the IAAFT surrogates are much wider than those associated with the shuffled surrogates. This can only be explained by assuming that IAAFT surrogates carry nonlinearities other than those induced by the data point distribution with a given the power spectrum. However, the three time se-ries can be attested to bear dynamic nonlinearities with a probability value of at least $\approx 10^{-5}$. While commonly used measures have very long computation times, the current results are verified by smooth null distributions because orders of magnitudes more surrogate realizations can be analyzed.

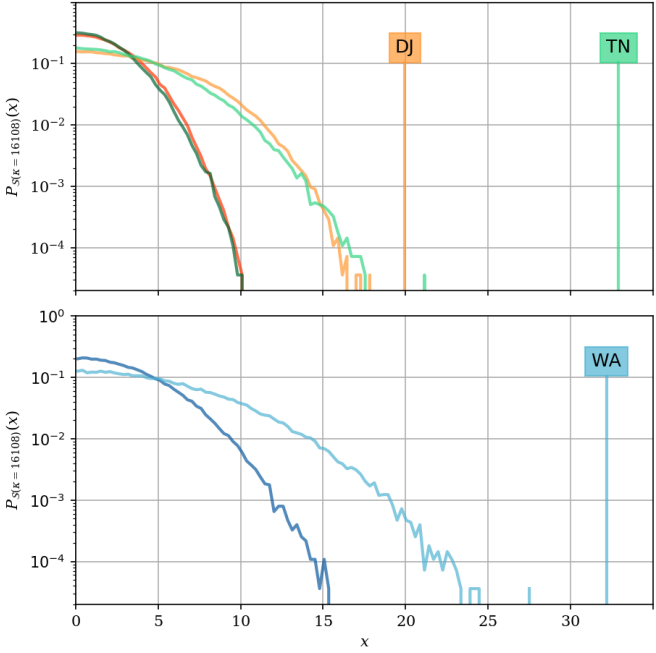

FIG. 7. Top: The distribution of the SRT results for the surrogates and the original data sets of DJ and the TN. Bottom: The distribution of the results for the WA data set and its surrogates. The distributions are grouped by their colors (orange for DJ, green for TN and blue for WA). The shuffled surrogates are darker than the IAAFT surrogates. The vertical lines labeled by the names indicate the results for the original data sets. IAAFT surrogates tend to produce much wider distributions than the shuffled time series.

### C. Comparison to other measures

In this section we compare the SRT to some well-established measures. Instead of plotting the result distributions, we specify a significance estimate by means of a modified z-score:[50]

$$\Sigma = \frac{0.6745 \cdot |r - \mathrm{med}(X)|}{\mathrm{med}(|X - \mathrm{med}(X)|)} \quad (11)$$

Here, med($\circ$) is the median of a distribution of random valiables $\circ$. $X$ is distribution of the test results of the surrogates, i.e. the null distribution. $r$ is the test result of the time series in question.

Three measures have been selected for comparison: a nonlinear prediction error based measure (NLPE),[5,6] a correlation integral (CI)[7,51] and the time reversibility (TR).[52]

1. NLPE: The idea behind the nonlinear prediction error (NLPE)[5,6] is to predict the trajectory of a chosen data point by averaging over the trajectories of its neighboring data points. If the prediction diverges significantly from the real path, chaotic dynamics are assumed to underlie the process.

To perform this analysis, one starts with embedding the time series into a higher dimensional phase space. Then, one data point $\mathbf{x}_t$ is selected and a set of corresponding neighbors is determined by either selecting all points in a spherical region around $\mathbf{x}_t$ (*fixed ball* or *soft ball*) or by taking a fixed number $N_{nn}$ of nearest neighbors $\mathbf{x}_{t_n}$ (*fixed mass*). In our analysis, the last option is chosen:

$$f_\tau(\mathbf{x}_t) = \frac{1}{N_{nn}} \sum_{n=0}^{N_{nn}-1} \mathbf{x}_{t_n+\tau} \quad (12)$$

The next step is to calculate the temporal mean of the squared distances between prediction and real trajectory over all times $t$. The NLPE finally becomes the square root of this mean:

$$\Psi(\tau) = \left\langle [f_\tau(\mathbf{x}_t) - \mathbf{x}_{t+\tau}]^2 \right\rangle_t^{\frac{1}{2}} \quad (13)$$

For the Dow-Jones data set, we use the embedding dimension $d = 8$ from a publication by Small and Tse[53] in which they studied the very same time series only over a different period of time. The number of nearest neighbors is $N_{nn} = d + 1 = 9$.[6] The maximal test significance is obtained by setting $\tau = 1$.

For WA, we select the parameters according to a Publication by Ragwitz and Kantz,[54] where also a time series of wind velocities has been analyzed: $d = 20$ and $N_{nn} = 50$. Here again, we obtain the maximal significance for $\tau = 1$.

2. CI: A set of fractal dimension estimates for time series has been introduced by Grassberger and Procaccia in 1983.[7,51] In their work the order-$q$ correlation sum is defined as

$$C_q(\epsilon) = \left\langle \langle \Theta(\epsilon - r_{ij}) \rangle_{i \neq j}^{q-1} \right\rangle_j \quad (14)$$

$$= \frac{1}{N} \sum_{j=0}^{N-1} \left[ \frac{1}{N-1} \sum_{i=0, i \neq j}^{N-1} \Theta(\epsilon - r_{ij}) \right]^{q-1}. \quad (15)$$

Here, $r_{ij}$ is the Euclidean distance between the embedded data points $\mathbf{x}_i$ and $\mathbf{x}_j$, $\Theta(\circ)$ is the Heaviside step function ($\Theta(\circ) = 1$ if $\circ > 0$ and $\Theta(\circ) = 0$ if $\circ < 0$), $\epsilon$ is a distance threshold. In this analysis, we evaluate $C_q(\epsilon)$ for $q = 2$ and values of $\epsilon$ that are again chosen to maximize the test significance: $\epsilon_{DJ} = 0.01$, $\epsilon_{TN} = 2.5$ and $\epsilon_{WA} = 1.0$. The embedding parameters are the same as in the NLPE setting.

3. TR: The time-reversibility[52] is defined as

$$\mathcal{T}(\tau) = \left\langle (x_{t+\tau} - x_t)^3 \right\rangle_t = 3 \left[ \left\langle x_{t+\tau} x_t^2 \right\rangle_t - \left\langle x_{t+\tau}^2 x_t \right\rangle_t \right]. \quad (16)$$

Significant deviations of $(T(\tau))$ from 0 indicate that the signal is not invariant under time reversal. Although this "is a sufficient and powerful indicator of nonlinearity", it is "not a necessary condition", as already mentioned by Schreiber and Schmitz.[52] For each evaluation we selected another $\tau$ that maximized the test result.

| Data | Surrogate | $\mathcal{S}(\kappa)$ | NLPE | CI | TR |
|------|-----------|------|------|------|------|
| DJ | shuffled | 14.8 | 42.2 | 81.3 | 10.9 |
| DJ | IAAFT | 8.4 | 41.0 | 74.5 | 14.4 |
| TN | shuffled | 24.2 | 51.0 | 70.3 | 24.1 |
| TN | IAAFT | 12.9 | 52.1 | 73.3 | 25.5 |
| WA | shuffled | 14.4 | 46.7 | 4082.7 | 24.2 |
| WA | IAAFT | 9.3 | 38.7 | 3399.4 | 26.0 |

TABLE I. Modified z-scores (Eq. (11)) for dynamic nonlinearities being present in the empirical time series as obtained by the SRT ($\mathcal{S}(\kappa)$), the nonlinear prediction error (NLPE), the correlation integral (CI) and the time reversibility (TR). Each result is based on 100 surrogate evaluations. $\mathcal{S}(\kappa)$ has been evaluated at $\kappa = 16108$.

Each method has been evaluated for both shuffled and IAAFT surrogates. The results are presented in Table I.

All test results are highly significant which makes a detailed comparison of the test methods obsolete. The smallest z-score of the SRT is 8.4. However, a noteworthy observation is that the significance of the SRT for IAAFT surrogates is systematically lower than for shuffled surrogates. The other measures do not show this pattern. As earlier studies have already shown,[20,55] the IAAFT surrogates show more phase correlations and hence more nonlinear artifacts than the shuffled surrogates (see also Fig. 7).

## IV. CONCLUSIONS AND OUTLOOK

In this Paper, we analytically characterized implications of leptokurtosis on the Fourier phases. As the latter must contain all nonlinear information of a time series, we proceeded with deriving a methodology, the SRT, that can capture anomalies among them. With it, we first confirmed the increase of nonlinearities with growing leptokurtosis and secondly tested empirical data for nonlinearities. From smooth null distributions representing SRT evaluations of a large number of IAAFT surrogates, we can infer p-values of at least $10^{-6}$ for the tested time series. Defining the null hypothesis by the shuffled surrogates, which is valid for the financial and tailored time series, even results in p-values many orders of magnitude higher.

While in this Paper we used the SRT only to detect nonlinearities in given data sets and to quantify the impact of static nonlinearities, it is also possible to apply scale dependent variants. I.e., the SRT can be determined for selected frequency bands. Further studies may use it to discover and quantify burstiness in even very noisy data sets. Phase walk analysis may therefore become a powerful alternative to other nonlinear measures, like, e.g., the nonlinear prediction error or correlation dimensions, with highly reduced computational effort.

While this manuscript is largely restricted to the analysis of leptokurtic data, future work may allow for unwinding other types of phase entanglement, helping to better understand Fourier phase information. This might not only be interesting from a theoretical point of view but can be of great value for financial applications, health sciences or even disaster prevention. If the phase information that is relevant to characterize a nonlinear time series can efficiently be parameterized by only a few values, very effective forecasting techniques, which are based on the essential Fourier phase information become conceivable. First ideas pointing in this direction can be found in a publication from 2015 by Räth et al..[21] Recent work suggests promising development in forecasting techniques,[56,57] but is still based on computationally intense algorithms and neglects the very habitat of nonlinearities, that is, the Fourier phases. In an often quoted article from 1999, Ivanov et al.[58] have shown that measuring a decline in a multifractal measure can indicate life-threatening heart conditions, but they also uncovered that the *"nonlinear properties of the healthy heart rate are encoded in the Fourier phases"*. This should be a motivation for further research on the meaning of Fourier phases, i.e. on the *nonlinear heart* of time series.

## ACKNOWLEDGEMENTS

This work has made use of data provided by the National Renewable Energy Laboratory ("NREL"), which is operated by the Alliance for Sustainable Energy, LLC ("ALLIANCE") for the U.S. Department Of Energy ("DOE") Furthermore, we would like thank Ingo Laut for fruitful discussions, that, not at least, lead to writing this Paper and for carefully reading the manuscript.